\documentstyle[twocolumn,floats,aps,prl,psfig]{revtex}
\begin{document}
\draft
\begin{title}
{Asymptotic scaling of the square lattice 
 quantum Heisenberg antiferromagnet}
\end{title}
\author{Jae-Kwon Kim}
\address {Center for Simulational Physics, \\
The University of Georgia, Athens, GA 30602}
\author{Matthias Troyer}
\address{Institute for Solid State Physics \\
University of Tokyo, Roppongi 7-22-1, Tokyo 106, Japan}
\maketitle
\begin{abstract}
We present thermodynamic measurements of various physical observables
of the two dimensional S=1/2 isotropic quantum Heisenberg
antiferromagnet on a square lattice, obtained by quantum Monte
Carlo. In particular we have been able to measure the infinite volume
limit of the uniform susceptibility up to the inverse temperature
$\beta=40$, the analysis of which reveals the correct asymptotic
behavior in excellent agreement with the prediction of chiral
perturbation theory. The issue of the existence of a crossover
from quantum critical to renormalized classical regime is clarified.
\end{abstract}
The square lattice quantum Heisenberg antiferromagnet (QHA) is important
in condensed matter physics since it describes the critical behavior of
undoped insulating parent compounds of the high $T_c$ superconductors.
Some materials believed to be well described by these models are
${\rm La}_2{\rm CuO}_4$ \cite{lacuo}, ${\rm Sr}_2{\rm CuO}_2{\rm Cl}$
\cite{GRE}, ${\rm La}_2{\rm NiO}_4$ \cite{NAK} and 
${\rm K}_2{\rm NiF}_4$ \cite{GRE}.  Most existing theoretical
treatments of the long wavelength, low energy behavior of the QHA
are actually based on an effective field theory, the (2+1)D $O(3)$
non-linear $\sigma$ model (NL$\sigma$M) \cite{CHA,HN,CHU}.

Although the mapping from the QHA to the NL$\sigma$M may be justified
on the general ground of universality, it is rigorous only for
sufficiently large values of the magnitude of the quantum spin
($S$). There is also a subtle problem due to the existence of Berry
phase terms, which are present in the effective field theory of the
QHA
\cite{berry}, but not in the NL$\sigma$M. This term is believed not to
be relevant \cite{CHU,TRO}, but this is still a matter of debate
\cite{CHA2}.

The results of previous experimental and numerical studies on the
S=1/2 QHA show a leading exponential temperature dependence of the
correlation length, in agreement with the prediction of the
theories \cite{CHA,HN,CHU}. There nevertheless exist some systematic
discrepancies between the observed values and theoretical predictions:
The two-loop order formula of the correlation length deviates by more
than 15\% from both measurements \cite{GRE,NAK} and numerical
simulations \cite{DING,ELS}.  The deviation becomes worse as the value
of $S$ increases \cite{ELS}, in contrast to naive expectations. Recent
neutron scattering experiments on $Sr_2CuO_2Cl_2$ \cite{GRE}, which is
regarded as a more ideal realization of the S=1/2 QHA than the
previously used compound $La_2CuO_4$ \cite{lacuo}, show systematic
deviations from the theoretical prediction for the peak value of the
static structure factor.  The presence of a crossover from quantum
critical to renormalized classical behavior predicted by the
theories \cite{CHA,CHU} remains controversial with differing claims
reported in the literature \cite{DING,SOK,GRE}.

In this Letter we present results of a very large scale quantum Monte
Carlo study on the S=1/2 square lattice QHA up to the linear size of
the lattice L=1000.  Our study resolves the above discrepancies, and
clearly confirms the validity of the theory \cite{CHA,HN,CHU}.  Using
the continuous time version \cite{BEARD1} of the loop algorithm
\cite{EVE} that eliminates the systematic error due to finite
Suzuki-Trotter number, we measure the thermodynamic values (infinite
volume limit values) of various physical observables such as the
uniform susceptibility ($\chi_u$), staggered susceptibility
($\chi_s$), second moment correlation length ($\xi$), peak value of
the staggered structure factor $S_{\bf Q}$ at ${\bf Q}=(\pi,\pi)$ and
internal energy (${\cal E}$).  For each $T$ and $L$ we performed
$10^6 \sim 10^7$ single loop updates after thermalization.

In order to monitor finite size effects in our measurements, we
repeated the measurements for each temperature with varying lattice
size and found that the measured values of $\xi$, $\chi_s$ and $S_{\bf
Q}$ become size independent under the condition $L/\xi \gtrsim 7$,
within the typical relative statistical errors of 0.3 percent or
better.  This condition is very restrictive for the measurements of
the infinite volume value of those quantities. It turns out however
that $\chi_u$ and ${\cal E}$ can be measured reliably on much smaller
lattices. The uniform susceptibility could be measured up to $\beta
\equiv J/T =40$, where $\xi$ is of order ${\cal O}(10^{19} \sim
10^{20})$. Varying L from 20 to 120 we found that the data are already
size independent for $L \ge 80$ within the typical statistical error
of 0.3 percent or better.

A selection of our data is shown in Table 1. The complete data
will be presented in a forthcoming publication \cite{fullpaper}.
In the following we discuss our measurements and fit them to 
theoretical predictions.
\\

{\it Uniform susceptibility}: 
Chiral perturbation theory \cite{HN},
for $\chi_u$ as $T \to 0$, predicts
(with convention $J = \hbar = g\mu_B = k_B=1$ hereafter)
\begin{equation}
\chi_u^{HN} = {2 \over 3} \chi_{\perp} \left[ 1 +{{T} \over {2\pi \rho_s}} 
		+ \left({{T} \over {2\pi \rho_s}}\right)^{2} \right ]
\label{eq:chiu_asy}
\end{equation}
where $\chi_{\perp} = \rho_s / c^2$. 
It turns out that the fit of our data to Eq. (\ref{eq:chiu_asy}) becomes
stable only for data with $ \beta \ge 4.5$, with $\chi^2/N_{DF} \simeq 0.6$. 
The estimated values of the parameters  from the fit are
\begin{equation}
\rho_s = 0.179(2),~~c = 1.652(1)  \label{eq:par_chiu}.
\end{equation}
These values are in remarkably good agreement with those obtained from 
spin wave theory of the QHA \cite{HAM}.

Our estimate of $\chi_{\perp}$ is
\begin{equation}
\chi_{\perp} = 0.0656(1),
\end{equation}
which is also consistent with  the result of the 
spin wave theory \cite{HAM}.

\begin{figure}
\begin{center}
\mbox{\psfig{figure=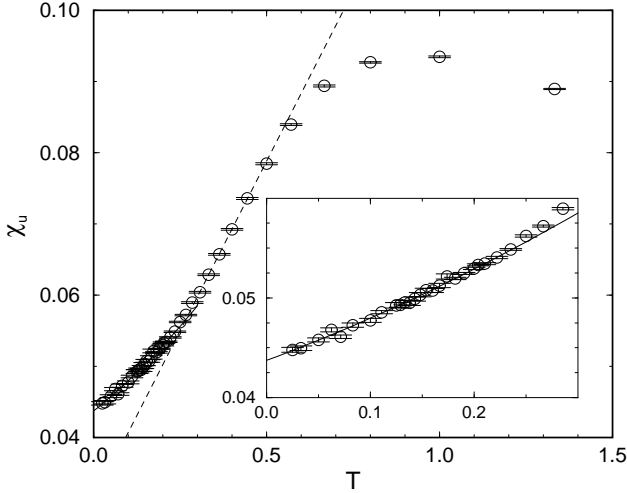,width=\hsize}}
\end{center}
\caption{The uniform susceptibility
$\chi_u$ versus $T$. The inset shows the low-$T$ region enlarged. The
fit to the low-$T$ prediction Eq. (\protect{\ref{eq:chiu_asy}}) is
drawn as a solid line, the fit to the quantum critical linear behavior
with is shown as a dashed line. In the latter fit the slope is
completely determined by the value of the $\chi_u(0)$, extracted from
the first fit.}
\label{fig:chiu}
\end{figure}

Also predicted theoretically \cite{CHU} is a crossover from quantum
critical behavior to renormalized classical one, that is, from \cite{CHU}
\begin{equation}
\label{eq:chi_qc}
\chi_u = {1\over c^2} \left[A_u T + B_u(\rho_s)\right]
\end{equation}
to Eq.(\ref{eq:chiu_asy}). In Eq. (\ref{eq:chi_qc}) the constant $A_u$
is universal. The best estimate from QMC is $A_u=0.26\pm0.01$
\cite{TRO}. $B_u$ is only known to leading order in a $1/N$ expansion
\cite{CHU} as $B_u\approx 0.57\rho_s$. We observe (Fig. 1) that in a
reasonably broad range $0.3 \lesssim T \lesssim 0.5$ $\chi_u$ is linear in $T$
with the expected slope $A_u$ and an offset $B_u\approx 0.47\rho_s$,
reasonably close to the analytical estimate.
\\

{\it Internal Energy}: We find that the energy ${\cal E}$ becomes size
independent under the condition $L/\xi \gtrsim 3$.  Our measurements are
over $0.25 \le \beta \le 5.5$, usually by varying $\beta$ by 0.25, on
lattices of size up to L=1000.  The theoretical prediction given in
Ref. \cite{HN} is
\begin{equation}
{\cal E}(T) = E_0 +  E_3 T^3 + E_5 T^5,~~E_3= 
{{2\zeta(3)} \over {\pi c^2}}.
\label{eq:int_eny}
\end{equation}
Due to considerable uncertainties in three parameter fits, we here fix
the value of $E_3$ with the value of $c$ given from the fit of
$\chi_u$, i.e, $c=1.652$ rather than treating $E_3$ as a fitting
parameter.  It turns out that only data for $\beta \gtrsim 4.25$ fit
reasonably well to Eq. (\ref{eq:int_eny}) with $\chi^2/N_{NF} \simeq
1.0$.  We observe however that the fit is still slightly unstable in
the sense that the values of the fitting parameters change mildly with
the range of $T$ selected for the fit. The ground state energy we have
extracted is
\begin{equation}
E_0 = -0.66953(4), \label{eq:par_ery}
\end{equation}
which may be compared with $E_0 = -0.6693 \sim -0.6694$
obtained from the ground state properties of QHA \cite{RUN,SAN}. 
Due to the mild increasing tendency of $E_0$ with restricting the
fit to lower temperatures, our estimate (Eq. (\ref{eq:par_ery})) 
may be regarded as a weak lower bound of the correct value.\\

{\it Correlation length}:
The prediction of $\xi$ from the two loop order of the 
chiral perturbation theory up to the first order of $T$ reads \cite{HN}
\begin{equation}
\xi_{HN} = {e \over 8}~~ {c \over {2\pi\rho_{s}} } \exp
\left({ {2\pi\rho_{s}} \over {T}} \right)
\times \left[ 1 - {{T} \over {4\pi\rho_{s}} } \right]  \label{eq:HN}
\end{equation}

\begin{figure}
\begin{center}
\mbox{\psfig{figure=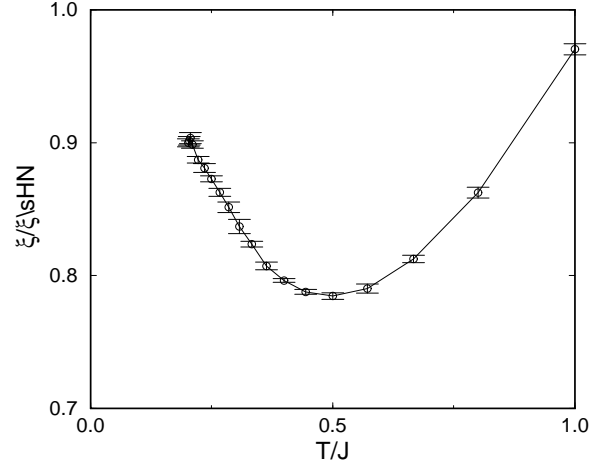,width=\hsize}}
\end{center}
\caption{Ratio of the measured and predicted correlation length, i.e.,
$\xi/\xi_{HN}(\rho_s,c)$ with the values of $\rho_s$
and $c$ given in Eq. (2), as a function of $T$.}
\label{fig:cl}
\end{figure}

We measured the infinite volume limit $\xi$ over the 
inverse temperature range
$0.25 \le \beta \le 4.95$, corresponding to $0.289(2) \le \xi \le
120.5(4)$, and fitted our data to Eq. (\ref{eq:HN}) in various
temperature ranges . We find that the fits are unstable, i.e., the
value of $\rho_s$ ($c$) decreases (increases) systematically as the
data at higher T are removed in the fit. Considering data with
$\xi_{min}= 39.2(1)$ in the fit we find
\begin{equation}
\rho_s = 0.185(1), ~~c = 1.442(3).  \label{eq:par_{cl}}
\end{equation}
Due to the systematic tendency this value of $\rho_s$ ($c$) should be
regarded as the upper (lower) bound of the correct asymptotic value.
Although those extracted from our $\chi_u$ data ,
Eq. (\ref{eq:par_chiu}), are indeed consistent with the the bounds,
our $\xi$ over $0.25 \le \beta \le 4.95$ strongly deviate from the
asymptotic expression ( Fig.\ref{fig:cl}). However we wish to note
that both our data and the experimental measurements \cite{GRE} can be
fitted quite well by Eq. (\ref{eq:HN}) with the correct $\rho_s$ if
one leaves the prefactor a free fitting parameter, as was done in
Ref. \cite{GRE}. 

The theory \cite{CHA,CHU} also predicts a crossover from the behavior
in the quantum critical regime, given by $c/\xi(T) = A_{QC}\rho_s +
B_{QC} T$, to the asymptotic behavior Eq. (\ref{eq:HN}).  The
expressions for $A_{QC}$ and $B_{QC}$ were obtained to one-loop order
of the renormalization group calculation \cite{CHA,CHU}.  A previous
Monte Carlo study \cite{DING} claimed that all $\xi$ fit a simple
exponential form even for $\beta$ as low as $0.25$,  contrary to our
data. They thus concluded that there is no crossover.  On the other
hand a series expansion study for $\xi\lesssim 10$ claimed to see a
crossover \cite{SOK}. We indeed observe that $1/\xi$ is linear in T
over $1.273(6) \le \xi \le 3.25$, but the measured values of $A_{QC}$
and $B_{QC}$ are not consistent with the theoretical predictions
\cite{CHA,CHU}. Moreover, we observe such a linearity even in the 2D classical
Heisenberg model where there should be no such crossover. Our data are
not consistent with the one-loop order equations for the crossover
regime \cite{CHA,CHU,CHA2} either. Thus our results seem incompatible
with the scenario of a crossover in $\xi$.
\\

{\it The peak value of the static structure factor}:
Theory \cite{CHA,CHU,KOP} predicts  for low $T$ 
\begin{equation}
{S_{\bf Q}(T) \over \xi^{2}(T)} = A 2\pi M^2
\left({T\over 2\pi \rho_s}\right)^2 \left(1-C {T\over 2 \pi\rho_s}\right),
\label{eq:sq}
\end{equation}
with a universal constant $A$. $M\approx 0.307$ \cite{SAN} is the
ground state magnetization. However the experimental data suggested
\cite{GRE,NAK} that $S_{\bf Q}(T) / \xi^2(T)$ is temperature
independent over the temperature range accessible in the experiment.

\begin{figure}
\begin{center}
\mbox{\psfig{figure=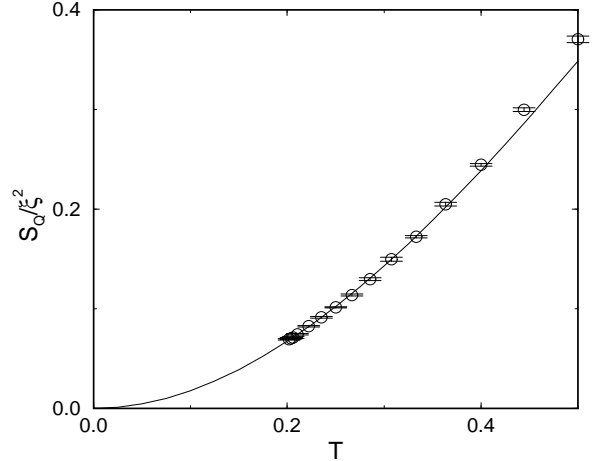,width=\hsize}}
\end{center}
\caption{Ratio of the staggered structure factor and the square of the
correlation length $S_{\bf Q}/\xi^2$ versus $T$.  The soild line is a
fit to the theoretical low-$T$ prediction Eq. (\protect{\ref{eq:sq}}).}
\label{fig:sq}
\end{figure}

We find that the data at $\xi\gtrsim 8.4$ fit with an acceptable
$\chi^2/N_{DF} = 0.9$ (see Fig. \ref{fig:sq}). The estimated value of
$C = -0.6(1)$ suggests that the effect of the correction term is less
than one percent for $T \lesssim 0.02$, which may be regarded as the
asymptotic regime for Eq. (\ref{eq:sq}). For the universal prefactor
we get the estimate $A\approx 4.0$. A series expansion study
\cite{SOK} got $A^s_{1/2}\approx3.2$ for spin $S=1/2$ and
$A^s_{\infty}\approx6.6$ for spin $S=\infty$. Our result clearly
shows, that as conjectured in Ref. \cite{SOK}, these values do not
agree because the models are not yet in the low-$T$ scaling regime.
As the fits are unstable in the sense that $A$ increases as we leave
out data at higher $T$ in the fit we view our estimate as a lower
bound.

We wish to note that for $\xi \ge 2.37(1)$ our data also fits to
$S_{\bf Q}(T)/\xi^2(T) \sim T^N$ with $N\simeq 1.85(3)$
with an acceptable value of $\chi^{2}/N_{DF}\simeq 0.75$.

Our data are definitely in agreement with the theory but not with the
analysis from experiment \cite{GRE}. The comparison with experimental
data will be discussed in more detail in a forthcoming publication
\cite{fullpaper}.
\\

{\it The staggered susceptibility}: In the classical high-T and in the
low-T renormalized classical regime the staggered susceptibility is
expected to be related to the staggered structure factor as
\cite{CHA,CHU}
\begin{equation}
{S_{\bf Q}\over T\chi_{ST}} = 1.
\end{equation}
\begin{figure}
\begin{center}
\mbox{\psfig{figure=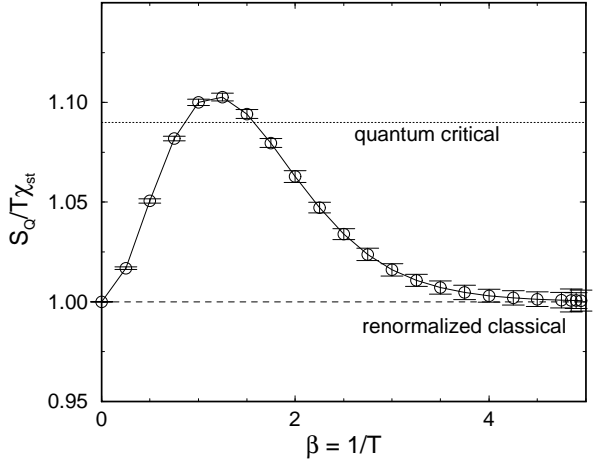,width=\hsize}}
\end{center}
\caption{Ratio of the staggered structure factor and susceptibility
$S_{\bf Q}\over T\chi_{ST}$ versus $\beta=J/T$. This ratio should be
$1$ (dashed line) both in the renormalized classical~($T\rightarrow 0$) and
high temperature regimes ($T\rightarrow\infty$). In the quantum critical regime
it is expected to be $1.09$.}
\label{fig:chist}
\end{figure}
This prediction agrees with QMC measurements by Sandvik {\it et
al.} \cite{SAN2}. Our data, shown in Fig. \ref{fig:chist} also confirm
it, with even higher precision. In the quantum critical regime this
ratio is expected to be also constant but with a value of
\cite{CHU,SOK}
\begin{equation}
{S_{\bf Q}\over T\chi_{ST}} \simeq 1.09.
\end{equation}
Instead of an expected plateau with this value the data however show a
peak with a maximum around $T\sim 0.8$ \cite{SAN2}. While the peak
value is close to the predicted quantum critical regime the
temperature is outside the range where quantum critical behavior is
observed in the uniform susceptibility. Thus we hesitate to ascribe
this maximum to quantum critical behavior.
\\
{\it Conclusion and discussion}: We believe that we have presented for
the first time a set of thermodynamic data which displays the
asymptotic behavior predicted by chiral perturbation theory. The
values of the spin-stiffness constant and spin wave velocity agree
well with those given by the spin wave theory of the QHA. This
agreement is truly remarkable since the chiral perturbation theory is
on the NL$\sigma$M whereas the spin wave theory is on the QHA. Our
results strongly confirm the validity of the map from the QHA to the
NL$\sigma$M in describing the long distance behavior of the former.
Our estimates of $\rho_s$ and $c$ from the analysis of $\chi_u$ may be
compared with previous estimates based on the size dependence formula
near the ground state, i.e., $\rho_s =0.185(2)$ and $c
=1.68(1)$ \cite{BEARD1}.  They are extracted, however, by three
parameter fits, which generally involve rather large uncertainties in
estimates.

The fact that the asymptote of $\chi_u$ manifests itself only at such
low temperatures may account for previous puzzles raised in studies of
the correlation length. The deviation of $\xi$ from Eq. (\ref{eq:HN})
is reduced to approximately 10 percent at $\beta\simeq 4.9$ from the
20 percent deviation seen previously at $\beta=2.5$.  In fact this is
similar to the 2D classical Heisenberg model as demonstrated by recent
numerical studies \cite{KIM,BUTE}. Our data of $S_{\bf Q}$ and $\chi_u$
also indicate that one needs to probe data with much lower $T$ for the
correction term to be safely ignored.

Another puzzle raised in experimental \cite{GRE,NAK} and numerical
\cite{SOK} studies was: why is the agreement of the measurements worse
for larger spins $S$? Equation (\ref{eq:HN}) depends on the value of
$S$ only implicitly through the $S$-dependence of $\rho_s$ and
$c$. The reason for the larger discrepancies at larger $S$ is implicit
in the assumptions in the theory: it is valid only for ${T\over
2\pi\rho_s}<<1$ and ${T\over 2\pi\rho_s}<<{c\over 2\pi\rho_s}\sim
1/S$. The latter condition restricts the validity of the theory to
larger and larger correlation lengths as the spin $S$ is increased.
      
We also clarify the issue of the existence of the crossover. As
suggested in Ref. \cite{CHU} the uniform susceptibility, where all
logarithmic corrections in the theory cancel, shows a clear crossover
from a quantum critical to the renormalized classical regime around
$T\sim J/3$. In all other quantities however no good evidence for a
crossover can be observed.

After completing our simulations, we became aware of a recent related
preprint \cite{BEARD2} that addresses similar subjects.  The authors
claim to extract thermodynamic $\xi$ up to $\beta =12$, with
corresponding $\xi$ of order of $10^5$, based on recently developed
finite size scaling extrapolation methods \cite{KIM}. This finite size
scaling method requires the existence of a universal scaling function
that has no explicit $T$ dependence. The existence of this function
can be explicitly checked numerically. However, contrary to the
corresponding classical case \cite{KIM}, we observed, in a
high-precision study, that in this quantum spin system the scaling
function becomes approximately $T$ independent only when $T$ is very
small. Detailed accounts of this issue will appear in a separate paper
\cite{KIM2}.

The simulations were performed on the 1024-node Hitachi SR2201
massively parallel computer of the computer center of the University
of Tokyo and used about 250 000 hours of CPU time. M.T. was supported
by the Japanese Society for the Promotion of Science.

\widetext

\begin{table}
\caption{A selection of thermodynamic data for the 2D QHA in the range 
         $0.25 \le \beta \le 4.95$.}
\label{table_bulk}
\begin{tabular}{ccc ccc ccc c}
$\beta$     &0.25     &1.75   &2.25        &2.75     &3.25       &3.75   
                                           &4.25     &4.75       &4.95\\ \hline
$\xi$       &0.289(2) &2.37(1)&4.46(1)     &8.38(1)  &15.7(1)    &29.0(1)
                                           &52.8(2)  &95.7(3)    &120.5(4) \\
$\chi_{st}$ &0.0811(2)&4.185(7)&12.54(3)   &38.2(1)  &117.3(3)   &357.5(9)
                                           &1082(3)  &3230(10)    &4980(17) \\
$\chi_u$   &0.0487(1)&0.0840(1)&0.0735(1)  &0.0656(1)&0.0604(1) &0.0572(1)
                                           &0.0550(1)&0.0535(1) &0.0531(1) \\
${\cal E}$ &-0.0986(1)&-0.5609(1)&-0.6162(1)&-0.6432(1)&-0.6559(1)&-0.6618(1)
					  &-0.6648(1)&-0.6663(1) &-0.6676(1) \\
$S_{\bf Q}$  &0.329(1) &2.63(1) &5.96(1)     &14.40(2) &36.91(5)   &95.7(2)
                                           &254.8(5) &682(2)     &1007(5) \\
\end{tabular}
\end{table}


\begin{thebibliography}{99}
\bibitem{lacuo} Y. Endoh {\it et al},
                Phys. Rev. B {\bf 37}, 7443 (1987);
                K. Yamada {\it et al}, Phys. Rev. B {\bf 40}, 4557 (1989);
\bibitem{GRE} M. Greven, R. J. Birgeneau, Y. Endoh, M. A. Kastner, M.
              Matsuda and G. Shirane, Z. Phys. B {\bf 96}, 465 (1995).
\bibitem{NAK} K. Nakajima, K. Yamada, S. Hosoya, Y. Endoh, M. Greven and
              R.J. Birgenau, Z. Phys. B {\bf 96}, 479 (1995)
\bibitem{CHA} S. Chakravarty, B. I. Halperin, and D. R. Nelson,
              Phys. Rev. B {\bf 39}, 2344 (1989)
\bibitem{HN} P. Hasenfratz and F. Niedermayer, Z. Phys. B {\bf 92}, 91 (1993)
\bibitem{CHU} A. V. Chubukov, S. Sachdev, and J. Ye, Phys. Rev. B {\bf 49},
               11919 (1994)
\bibitem{berry}  F. D. M. Haldane, Phys. Rev. Lett. {\bf 61}, 1029 (1988).
\bibitem{TRO} M. Troyer {\it et al.}, cond-mat/9702077, J. Phys. Soc. Jpn.,
  in press; M. Troyer and M. Imada in {\it Computer Simulations in Condensed
  Matter Physics X}, ed. D.P. Landau {\it et al.,} (Springer Verlag,
  Heidelberg, 1997).
\bibitem{CHA2} S. Chakravarty in {\it Random magnetism and high
temperature superconductivity}, ed. by W.P. Beyermann, N.L. Huang-Liu
and D.E. MacLaughlin, World Scientific (Singapore 1993).
\bibitem{DING} M. S. Makivic and H.-Q. Ding, Phys. Rev. B {\bf 43},
	       3562 (1991)
\bibitem{ELS} N. Elstner {\it et al}, Phys. Rev. Lett. {\bf 75}, 938 (1995)
\bibitem{SOK} A. Sokol, R. L. Glenister, and R. L. Singh, Phys. Rev. Lett.
               {\bf 72}, 1549 (1994)
\bibitem{BEARD1} B. B. Beard and U.-J. Wiese, Phys. Rev. Lett. {\bf 77},
                5130 (1996)
\bibitem{EVE} H. G. Evertz, M. Marcu, and G. Lana, Phys. Rev. Lett. {\bf 70},
               875 (1993)
\bibitem{fullpaper} M. Troyer and J.-K. Kim, in preparation.
\bibitem{HAM} J. Igarashi, Phys. Rev. B {\bf 46}, 10763 (1992);
              C. J. Hamer, Z. Weihong, and J. Oitmaa, Phys. Rev. B
	        {\bf 50}, 6877 (1994)
\bibitem{RUN} K. J. Runge, Phys. Rev. B {\bf 45}, 12292 (1992).
\bibitem{SAN} A. W. Sandvik, preprint, cond-mat/9707123
\bibitem{KOP}  P. Kopietz, Phys. Rev. Lett. {\bf 64}, 2587 (1990)
\bibitem{SAN2} A. W. Sandvik, A. V. Chubukov and S. Sachdev,
	       Phys. Rev. B {\bf 51}, 16483 (1995); A. W. Sandvik and 
               D. J. Scalapino, Phys. Rev. B {\bf 53}, R526 (1996).
\bibitem{KIM}  J.-K. Kim, Phys. Rev. D {\bf 50}, 4663 (1994);
S. Caracciolo, R. G. Edwards, A. Pellisseto, and A. D. Sokal,
                Phys. Rev. Lett. {\bf 75}, 1891 (1995).
\bibitem{BUTE} P. Butera and M. Comi, Phys. Rev. B {\bf 54}, 15828 (1996)
\bibitem{BEARD2} B.B. Beard, R. J. Birgeneau, M. Greven, and U.-J. Wiese,
		 preprint, cond-mat/9709110
\bibitem{KIM2} J.-K. Kim and M. Troyer, in preparation
\end{thebibliography}
\end{document}